# Emirati Speaker Verification Based on HMM1s, HMM2s, and HMM3s

Shahin, Ismail
Department of Electrical and Computer Engineering, University of Sharjah, Sharjah, United Arab Emirates
E-mail: ismail@sharjah.ac.ae

*Abstract* - **This work focuses on Emirati speaker verification systems in neutral talking environments based on each of First-Order Hidden Markov Models (HMM1s), Second-Order Hidden Markov Models (HMM2s), and Third-Order Hidden Markov Models (HMM3s) as classifiers. These systems have been evaluated on our collected Emirati speech database which is comprised of 25 male and 25 female Emirati speakers using Mel-Frequency Cepstral Coefficients (MFCCs) as extracted features. Our results show that HMM3s outperform each of HMM1s and HMM2s for a text-independent Emirati speaker verification. The obtained results based on HMM3s are close to those achieved in subjective assessment by human listeners.**

## I. INTRODUCTION

Speaker recognition branches into two different major branches: speaker identification and speaker verification (authentication). Speaker identification can be defined as the method of automatically deciding who is speaking from a group of given speakers. Speaker verification can be defined as the method of automatically admitting or refusing the identity of the claimed speaker. Speaker identification appears in criminal investigations to determine the suspected persons who uttered a voice captured at the incident of a crime. Speaker verification is widely used in security access to services via a telephone, including home shopping, home banking transactions using a telephone network, security control for private information areas, remote access to computers, and many telecommunication services [1]. Speaker recognition is categorized, based on the text to be spoken, into text-dependent and text-independent cases. In the text-dependent case, speaker recognition requires the speaker to utter speech for the same text in both training and testing phases, while in the text-independent case, speaker recognition is independent on the text being spoken.

## II. RELATED WORK

The vast majority of studies in speaker recognition area focus on speech spoken in English language [1-4]; on the other hand, very limited number of work focus on such area on speech uttered in Arabic language [5-9]. This is because of the limited number of available Arabic speech databases in this area [10-11]. In the Arab countries, Arabs communicate in one of the regional dialects, of which there are four main kinds: Egyptian, Levantine, North African, and Gulf Arabic (*e.g.* Emirati). To the best of our knowledge, this work is considered as the first work that has been carried out on Emirati (United Arab Emirates) speaker verification.

Alkanhal *et al* [5] implemented speaker verification using a Saudi-accented Arabic telephone speech database with 1033 speakers based on Gaussian Mixture Models (GMMs). Their study concluded that speaker verification system can be enhanced by merging scores of several utterances [5]. Alsulaiman *et al* [6] investigated Arabic speaker recognition using an openly accessible speech database called Babylon Levantine which is available from the Linguistic Data Consortium (LDC). In their investigation, they used Hidden Markov Models (HMMs) as classifiers. Both the parameters of HMM models and the size of speech features have been considered to study their impact on the recognition performance [6]. Their results showed that the recognition performance increases with the increase in the number of mixtures, until it reaches a saturation level which depends on the data size and the number of HMM states. Alarifi *et al* [7] proposed a text-dependent speaker verification system for Arabic language. They [7] used discrete representation of speech signals in terms of Mel-frequency cepstral coefficients (MFCCs), linear predictive coding (LPC), and perceptual linear prediction (PLP) as input features for the system and they employed Support Vector Machines (SVMs) as classifiers to verify the claimed speaker. In another study by Alarifi *et al* [8], they proposed a new Arabic text-dependent speaker verification system for mobile devices based on Artificial Neural Networks (ANNs) to verify the legal user and unlock his/her mobile device. Shahin and Ba-Hutair [9] focused in one of their studies on Emirati speaker identification systems in neutral talking environments based on each of Vector Quantization (VQ), GMMs, and HMMs as classifiers. The Emirati speech database was comprised of 25 male and 25 female Emirati speakers. There [9] results showed that VQ outperforms each of GMMs and HMMs for both text-dependent and text-independent Emarati speaker identification.

The focus of this research is on evaluating a text-independent speaker verification using Emirati speech database collected in neutral talking environments. The database was captured from 25 male and 25 female Emirati speakers who uttered 8 commonly used Emirati sentences. In this work, three different classifiers have been employed. These classifiers are: First-Order Hidden Markov Models (HMM1s), Second-Order Hidden Markov Models (HMM2s), and Third-Order Hidden Markov Models (HMM3s).

The remaining of the paper is structured as follows: Brief overview of HMM1s, HMM2s, and HMM3s is given in Section 3. Section 4 explains the captured speech database used in this work and the extracted features. Speaker verification algorithm based on HMM1s, HMM2s, and HMM3s and the experiments are discussed in Section 5. Decision threshold is presented in Section 6. Section 7 gives the results attained in the current work and their discussion. Finally, concluding remarks are given in Section 8.

## III. BRIEF OVERVIEW OF HMM1s, HMM2s, AND HMM3s

### A. HMM1s

HMMs can be expressed as being in one of the $N$ different states: 1, 2, 3,…, $N$, at any discrete time instant $t$. The given states are symbolized as,

$$s = \{s_1, s_2, s_3, \ldots s_N\}$$

which are originators of a state sequence $q_t$, where at any time $t$: q = {$q_1, q_2, \ldots, q_T$}. At any discrete time $t$, the model is in a state $q_t$. At the discrete time $t$, the model makes an arbitrary move to a state $q_{t+1}$. The state transition probability matrix $A$ decides the probability of the next transition between states [12], [13],

$$\mathbf{A} = [\,a_{ij}\,] \qquad i, j = 1, 2, \ldots, N$$

where $a_{ij}$ indicates the transition probability from a state $i$ to a state $j$.

In such models, the state sequence is a first-order Markov chain where the stochastic process is modeled in a 2-D matrix of a priori transition probabilities ($a_{ij}$) between states $s_i$ and $s_j$ where $a_{ij}$ are given as:

$$a_{ij} = \text{Prob}\left(q_t = s_j | q_{t-1} = s_i\right) \quad (1)$$

In such acoustic models, it is assumed that the state-transition probability at time $t+1$ depends only on the state of the Markov chain at time $t$. Readers can get more information about HMM1s from references [12], [13].

### B. HMM2s

The state sequence in HMM2s is a second-order Markov chain where the stochastic process is expressed by a 3-D matrix ($a_{ijk}$). Therefore, the transition probabilities in HMM2s are given as [14], [15]:

$$a_{ijk} = \text{Prob}\left(q_t = s_k | q_{t-1} = s_j, q_{t-2} = s_i\right) \quad (2)$$

with the constraints,

$$\sum_{k=1}^{N} a_{ijk} = 1 \qquad N \geq i, j \geq 1$$

The state-transition probability in HMM2s at time $t+1$ relies on the states of the Markov chain at times $t$ and $t-1$. More information about HMM2s can be obtained from references [14], [15].

### C. HMM3s

The underlying state sequence in HMM3s is a third-order Markov chain where the stochastic process is defined by a 4-D matrix ($a_{ijkw}$). Therefore, the transition probabilities in HMM3s are given as [16], [17],

$$a_{ijkw} = \text{Prob}\left(q_t = s_w | q_{t-1} = s_k, q_{t-2} = s_j, q_{t-3} = s_i\right) \quad (3)$$

with the constraints,

$$\sum_{w=1}^{N} a_{ijkw} = 1 \qquad N \geq i, j, k \geq 1$$

The probability of the state sequence, $Q \triangleq q_1, q_2, \ldots, q_T$, is defined as:

$$\text{Prob}(Q) = \Psi_{q_1} a_{q_1 q_2 q_3} \prod_{t=4}^{T} a_{q_{t-3} q_{t-2} q_{t-1} q_t} \quad (4)$$

where $\Psi_i$ is the probability of a state $s_i$ at time $t = 1$, and $a_{ijk}$ is the probability of the transition from a state $s_i$ to a state $s_k$ at time $t = 3$.

Given a sequence of observed vectors, $O \triangleq O_1, O_2, \ldots, O_T$, the joint state-output probability is expressed as:

$$\text{Prob}(Q, O | \lambda) = \Psi_{q_1} b_{q_1}(O_1) a_{q_1 q_2 q_3} b_{q_3}(O_3) \cdot \prod_{t=4}^{T} a_{q_{t-3} q_{t-2} q_{t-1} q_t} b_{q_t}(O_t) \quad (5)$$

Interested readers in HMM3s can obtain more information about HMM3s from references [16], [17].

## IV. SPEECH DATABASE AND EXTRACTION OF FEATURES

### A. Collected Speech Database

The captured speech database was comprised of 25 male and 25 female Emirati speakers spanning from the age of 14 to 27 year old. Each speaker uttered 8 common Emirati sentences that are commonly used in the United Arab Emirates society. The eight sentences were neutrally portrayed by each speaker (no stress or emotion) 9 times with a range of 1 – 3 seconds. These speakers were inexperienced to keep away from overstated expressions. The total collected number of utterances was 3600 (50 speakers × 8 sentences × 9 repetitions/sentence). The eight sentences are tabulated in Table I (the right column gives the sentences in Emirati accent while the left column gives the English translation),

The recorded database was captured in an uncontaminated environment in one of the studios in the College of Communication at the University of Sharjah in the United Arab Emirates by a speech acquisition board using a 16-bit linear coding A/D converter and sampled at a sampling rate of 44.6 kHz. The speech signals were then down sampled to 12 kHz. The signal samples were preemphasized and then segmented into frames of 20 ms each with 31.25% overlap between successive frames.

### B. Extraction of Features

Mel-Frequency Cepstral Coefficients (MFCCs) have been used in this study as the appropriate features that express the phonetic content of Emirati speech signals. These features have been broadly used in many fields of speech. Such fields are speech and speaker recognition. These coefficients have proven to outperform other coefficients in the two fields and they have shown to provide a high-level approximation of human auditory perception [18], [19]. In this research, a 38-dimension feature analysis of MFCCs was used to shape the observation vectors in each of HMM1s, HMM2s, and HMM3s. A continuous mixture observation density was selected for each model. The number of states, $N$, in each model was 4 and the number of mixture components, $M$, was 32 per state.

## V. SPEAKER VERIFICATION ALGORITHM BASED ON HMM1s, HMM2s, AND HMM3s AND THE EXPERIMENTS

In the training phase in each of HMM1s, HMM2s, and HMM3s (three different and separate sessions), the $v$th speaker has been represented by a $v$th model. The $v$th model has been derived using the first four sentences with a repetition of nine utterances per sentence of the database. This gives a total of 36 utterances (4 sentences × 9 repetitions) for every speaker model.

In the evaluation phase in each of HMM1s, HMM2s, and HMM3s (three distinct and independent sessions), each one of the fifty speakers used nine utterances per sentence of the last four sentences (text-independent) of the database. The total number of utterances used in this phase is 1800 (50 speakers × 4 sentences × 9 utterances / sentence). In this work, 40 speakers (20 speakers per gender) are used as claimants and the rest of the speakers are used as imposters.

To verify the speaker identity based separately on each of HMM1s, HMM2s, and HMM3s, the log-likelihood ratio in the log domain has been computed as given in the following formula [20],

$$\Lambda_{model}(O) = log\left[P\left(O|\lambda_{model,C}\right)\right] - log\left[P\left(O|\lambda_{model,\overline{C}}\right)\right] \quad (6)$$

where, $\Lambda_{model}(O)$ is the log-likelihood ratio in the $log$ domain, $P(O|\lambda_{model,C})$ is the probability of the observation sequence $O$ given it comes from the claimed speaker, $P(O|\lambda_{model,\overline{C}})$ is the probability of the observation sequence $O$ given it does not come from the claimed speaker, and "$model$" represents HMM1s, or HMM2s, or HMM3s.

The probability of the observation sequence $O$ given it comes from the claimed speaker can be computed as [20],

$$log\ P(O|\lambda_{model,C}) = \frac{1}{T}\sum_{t=1}^{T} log\ P(o_t|\lambda_{model,C}) \quad (7)$$

where, $O = o_1 o_2 ... o_t ... o_T$ and $T$ is the utterance duration.

The probability of the observation sequence $O$ given it does not come from the claimed speaker can be computed using a set of $B$ imposter speaker models: $\{\lambda_{model,\overline{C}_1}, \lambda_{model,\overline{C}_2}, ..., \lambda_{model,\overline{C}_B}\}$ as,

$$log\ P(O|\lambda_{model,\overline{C}}) = \left\{\frac{1}{B}\sum_{b=1}^{B} log\left[P(O|\lambda_{model,\overline{C}_b})\right]\right\} \quad (8)$$

where $P(O|\lambda_{model,\overline{C}_b})$ can be computed using Eq. (7). Fig. 1 demonstrates a block diagram of speaker verification system.

## VI. DECISION THRESHOLD

In speaker verification problem, two kinds of error can take place. The two kinds are false rejection and false acceptance. When a legitimate identity claim is refused, it is called a false rejection error; in contrast, when the identity claim from an imposter is admitted, it is named a false acceptance error.

Speaker verification problem necessitates making a binary decision based on two hypotheses: Hypothesis $H_0$ if the observation sequence $O$ given it comes from the claimed speaker or hypothesis $H_1$ if the observation sequence $O$ given it does not come from the claimed speaker.

To admit or refuse the claimed speaker, a comparison between the log-likelihood ratio and the threshold ($\theta$) should take place as the last step in the verification procedure, *i.e.,* [20]

Admit the claimed speaker if $\Lambda(O) \geq \theta$

Refuse the claimed speaker if $\Lambda(O) < \theta$

Open set speaker verification often uses thresholding to make a decision if a speaker is out of the set. Both types of error in speaker verification problem depend on the threshold used in the decision making process. A strict value of threshold makes it hard for false speakers to be falsely admitted but at the expenditure of falsely refusing true speakers. In contrast, a relaxed value of threshold eases true speakers to be constantly admitted at the expenditure of falsely admitting false speakers. To place a proper value of threshold that meets with a desired level of a true speaker rejection and a false speaker acceptance, it is important to know the distribution of true speaker and false speaker scores. An acceptable process for setting a value of threshold is to start with a loose initial value of threshold and then let it adjust by setting it to the average of up-to-date trial scores. This loose value of threshold yields insufficient protection against false speaker trials.

## VII. RESULTS AND DISCUSSION

This work focuses on evaluating speaker verification using an Emirati speech database collected in neutral talking environments from 25 male and 25 female Emirati speakers uttering eight widely used sentences in the United Arab Emirates society. MFCCs have been used as the extracted features of the captured database. Three distinct classifiers have been employed to assess Emirati speaker verification systems. These classifiers are HMM1s, HMM2s, and HMM3s.

Table II demonstrates an Equal Error Rate (EER) percentage of a text-independent speaker verification using the Emirati database based on each of HMM1s, HMM2s, and HMM3s. This table apparently shows that HMM3s are superior to each of HMM1s and HMM2s for Emirati speaker verification.

A statistical significance test has been performed to show whether EER differences (EER based on HMM3s and that based on each of HMM1s and HMM2s) are real or simply due to statistical fluctuations. The statistical significance test has been carried out based on the Student's $t$ Distribution test as given by the following formula,

$$t_{1,2} = \frac{\overline{x}_1 - \overline{x}_2}{SD_{pooled}} \qquad (9)$$

where $\overline{x}_1$ is the mean of the first sample of size $n$, $\overline{x}_2$ is the mean of the second sample of the same size, and $SD_{pooled}$ is the pooled standard deviation of the two samples given as,

$$SD_{pooled} = \sqrt{\frac{SD_1^2 + SD_2^2}{2}} \qquad (10)$$

where $SD_1$ is the standard deviation of the first sample of size $n$ and $SD_2$ is the standard deviation of the second sample of the same size.

Using the last two equations and the results of Table II, the calculated $t$ values between HMM3s and each of HMM1s and HMM2s of EER are given in Table III. Table III illustrates that each calculated $t$ value is greater than the tabulated critical value at *0.05* significant level $t_{0.05} = 1.645$. Hence, it can be concluded from this experiment that Emirati speaker verification based on HMM3s outperforms that based on each of HMM1s and HMM2s.

Fig. 2 shows Detection Error Trade-off (DET) curve. This curve compares speaker verification using the Emirati speech database based on each of HMM1s, HMM2s, and HMM3s. This figure evidently demonstrates that HMM3s are superior to each of HMM1s and HMM2s for speaker verification using Emirati speech database in neutral talking environments.

The achieved results of EER percentage of a text-independent speaker verification using the Emirati database have been compared with those based on the state-of-the-art models and classifiers. Table IV shows EER percentage of a text-independent speaker verification using the Emirati database based on SVMs [21], GMMs [20], and VQ [22]. Tables II and IV apparently show that HMM3s are superior to each of HMM1s, HMM2s, SVMs, GMMs, and VQ for Emirati speaker verification.

An informal subjective assessment of HMM3s using the Emirati speech database has been conducted with ten nonprofessional Emirati listeners (human judges). A total of 400 utterances (50 speakers × 8 sentences) have been used in this assessment. During such an evaluation, each listener was independently asked to verify the claimed speaker for every test utterance. The average speaker verification performance based on the subjective assessment is 92.4%. This verification performance is close to that achieved in our current work using the same database based on HMM3s.

## VIII. CONCLUDING REMARKS

In this work, we evaluate a text-independent speaker verification using Emirati speech database in neutral talking environments based on each of HMM1s, HMM2s, and HMM3s. To the best of our knowledge, this is the first work conducted on speaker verification that has been tested on Emirati speech database. Our results show that HMM3s is superior to each of HMM1s and HMM2s for a text-independent Emirati speaker verification. An informal subjective assessment has been performed in this work. This assessment shows that speaker verification performance based on the subjective assessment is close to that attained in the current work using the same database based on HMM3s. Our plan in the near future will be focused on collecting Emirati speech database in each of stressful and emotional talking environments to study and assess speaker verification in these talking environments.

## ACKNOWLEDGEMENTS

The author wishes to thank University of Sharjah for funding this work through the competitive research project entitled "Capturing, Studying, and Analyzing Arabic Emirati-Accented Speech Database in Stressful and Emotional Talking Environments for Different Applications", No. 1602040349-P. The author also wishes to thank engineers Ahmad Ibrahim Al Hosani and Ahmed Esam Abdulqader at the University of Sharjah for capturing the Emirati speech database.

Table I. Emirati Speech Database and its English Translation

| No. | English Translation | Emirati Accent |
|---|---|---|
| 1. | Hello, how are you. | مرحبا الساع شحالك. |
| 2. | Happy fest and best wishes. | مباركن عيدكم وعساكم من عواده. |
| 3. | Congratulation for the new born baby and brought up in your glory. | مبروك ما ياكم ويتربى بعزكم. |
| 4. | The beauty of the dress is to be embroidered from origin. | حلات الثوب رقعته منه وفيه. |
| 5. | I wish to have threed on the breakfast. | خاطري في ثريد عالفطور. |
| 6. | Who is sitting next to you. | منو يالس حذالك. |
| 7. | We put the teapot on the fire and wait for the tea. | حاطين الغوري عالضو ونتريه الجاي يستوي. |
| 8. | Excuse me, a circumstance happened to me yesterday and I was not able to talk to you. | اسمح لي امس صار لي ظرف وما رمت ارمسك. |

Table II. EER Percentage of a Text-Independent Speaker Verification using Emirati Speech Database based on each of HMM1s, HMM2s, and HMM3s

| Classifier | EER % |
|---|---|
| HMM1s | 11.5 |
| HMM2s | 9.6 |
| HMM3s | 4.9 |

Table III. Calculated $t$ Values between HMM3s and each of HMM1s and HMM2s using Emirati Speech Database

| $t$ Value ($t_{1,2}$) | Calculated $t$ Value |
|---|---|
| $t_{HMM3s, HMM1s}$ | 1.703 |
| $t_{HMM3s, HMM2s}$ | 1.795 |

Table IV. EER Percentage of a Text-Independent Speaker Verification using Emirati Speech Database based on each of SVMs, GMMs, and VQ

| Classifier | EER % |
|---|---|
| SVMs | 7.3 |
| GMMs | 9.8 |
| VQ | 12.6 |

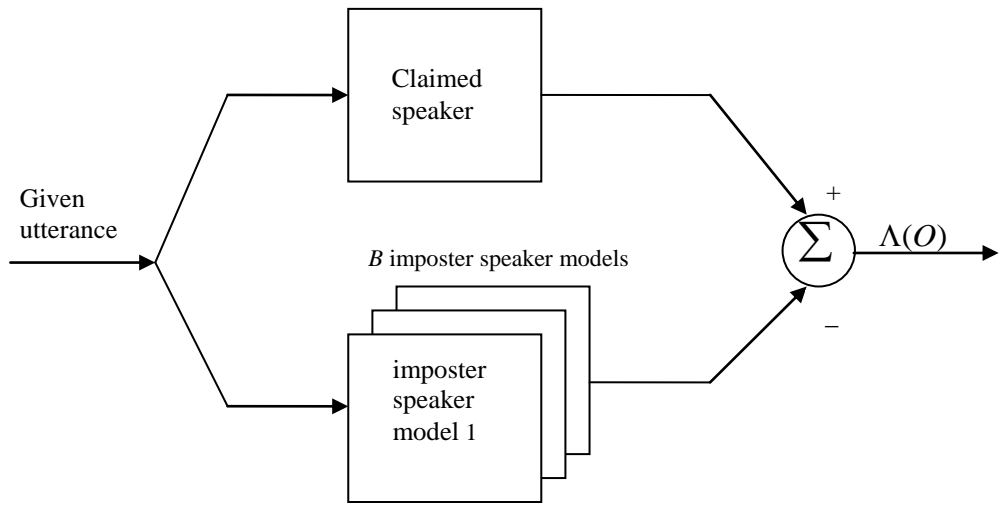

Fig. 1. Block diagram of speaker verification system

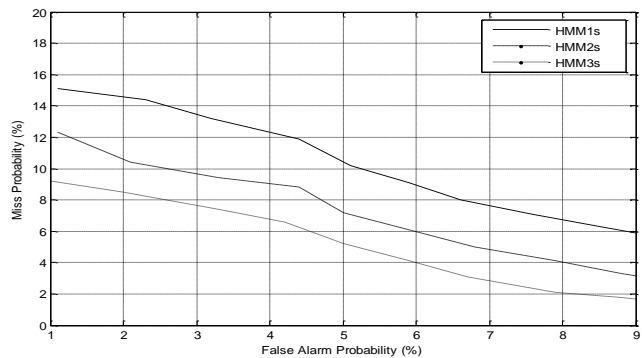

Fig. 2. DET curve based on each of HMM1s, HMM2s, and HMM3s using Emirati speech database